# The uniqueness of physical and chemical natures of graphene: their coherence and conflicts

E.F. Sheka

Peoples' Friendship University of Russia, 117198 Moscow, Russia

Molecular-crystalline duality of graphene ensures a tight alliance of its physical and chemical natures, each of which is unique in its own way. The paper examines the physical-chemical harmony and/or confrontation in terms of the molecular theory of graphene. Chemistry that is consistent with graphene physics expectations involves: small mass of carbon atoms, which provides a lightweight material; $sp^2$ configuration of the atoms valence electrons, ensuring a flat 2D structure of condensed benzenoid units; high strength of C-C valence bonds responsible for exclusive mechanical strength. Chemistry that is in conflict with graphene physics expectations covers: radical character of graphene material; collective character of electronic system of graphene, preventing from localization of its response on any external impact; the molecular nature and topochemical character of graphene mechanics; the molecular nature of graphene magnetism. Each of the properties is a direct consequence of the odd electron correlation, on one hand, and ensures a high instability of graphene material, on the other. Electron correlation inhibition by deposition of graphene monolayers on substrates is seen as a promising way to solve the problem.

1. **Introduction**

The nickname of a 'miracle material' has been widely accepted by the graphene community. It normally means the material superior properties. However, all the properties are only the outward manifestation of the graphene wonderful nature. The real miracle of graphene is that it is a union of two entities: the physical and chemical ones, each of which is unique in its own way. The physical superior properties of graphene are widely discussed. Much less has been told about its chemical uniqueness that is generated by carbon atoms packing in a flat honeycomb structure. The structure, based on benzenoid units, offers three neighbors to each carbon atom leaving the atom fourth valence electron on its own. These electrons form the pool of *odd* electrons whose behavior might change from the covalent bonding, characteristic for π electrons, to free electrons of radicals when the interaction between the electrons becomes weaker and weaker. The two electron states belong to different limit cases in terms of the electron correlation: π electrons, occupying the same place in the space, are not correlated while radical electrons are strongly correlated so that two electrons with different spins are space separated. π-Electrons of the benzene molecule belong to the first limit case while odd electrons of benzenoid units of graphene (as well as fullerenes and carbon nanotubes) are correlated, which is caused by the difference in the C-C bond lengths while the critical length of 1.395Å and below marks the first limit case [1-3].

Physics and Chemistry of graphene as any other outstanding organisms in any community can both strengthen, on one hand, and conflict each other, on the other. The graphene chemistry, which is consistent with physical expectations, covers the following issues.

1. Small mass of carbon atoms that provides the lightest material under ambient conditions;
2. $sp^2$ Configuration of the atom valence electrons that ensures a perfect flat 2D structure of condensed benzenoid units;
3. High strength of C-C valence bonds that is responsible for the exclusive mechanical strength.

If the first two issues need no comments, the latter deserves some explanation. Not long ago the presentation of carbyne as the stiffest carbon material has been heralded [4]. Presented results are based on an extensive computational study. A reconstruction of the equivalent continuum-elasticity representation was performed to determine Yang's moduli that served as a basis for approval of exceptional stiffness. This is a typical physical approach to the consideration of mechanical properties of graphene and related materials. However, basing on chemical concepts, the conclusion about the exceptional strength of the molecules with triple C-C bonds could be obtained practically without any computations. The fact is that since any elastic mechanical deformation is closely related to the elastic vibrations of the atoms, the main mechanical characteristics are tightly connected with parameters of the vibrational motion. For bodies with chemical bonds, to which nanocarbons belong, the elastic part of the tensile behavior, which is provided by stretching of these bonds, is obviously connected with the relevant force constants, or squared frequencies of stretching vibrations [5]. In view of this, elastic parameters such as Young's modulus and stiffness are proportional to squared frequencies of the relevant harmonic vibrations. Following this, a very simple way to compare Young's moduli of covalent bodies with either single or double and triple C-C bonds can be suggested. Thus, taking the Yang modulus of graphene, $Y_{C=C}$ as the reference, the following simple relations for $Y_{C-C}$ and $Y_{C\equiv C}$ moduli can be obtained

$$Y_{C-C} = (\nu_{C-C}/\nu_{C=C})^2 * Y_{C=C} \approx (0.5 \div 0.6) Y_{C=C} \qquad (1)$$

$$Y_{C\equiv C} = (\nu_{C\equiv C}/\nu_{C=C})^2 * Y_{C-C} \approx (1.8 \div 1.9) Y_{C=C} \qquad (2)$$

Numerical factors correspond to stretching vibrations of the corresponding C-C bonds, for which are taken vibrational frequencies related to ethane (~1100 cm$^{-1}$), ethylene (~1600 cm$^{-1}$) and acetylene (~2200 cm$^{-1}$) molecules, as average values. The factors are well consistent with reduced Yang modulus of graphane and fluorographene [6, 7] and twice enhanced modulus of carbyne [4].

The graphene chemistry, which is inconsistent with physics expectations, constitutes a longer list that includes but a few the most important issues, such as
1. The radical character of the graphene substance;
2. The collective behavior of the graphene odd electrons;
3. The molecular nature and topochemical character of graphene mechanics;
4. The molecular nature of graphene magnetism.

Each of the properties is a direct consequence of the odd electron correlation [2, 3]. The first issue provides the enhanced chemical reactivity of graphene. The second prevents localization of its response on any external impact. The next concerns a high sensitivity of graphene properties to external stress. The fourth makes the property size and shape dependent. A high propensity to $sp^2 \rightarrow sp^3$ transformation of carbon atom valence electrons, which violates flat 2D structure of a carpet of condensed benzenoid units, should be added to the set.

The very existence of objective obstacles to the successful execution of the promising graphene applications according to, for example, the 'Flagship graphene' program has been clearly evident. The standpoint of the situation was distinctly formulated by K.Novoselov et al

[8] in terms of the 'low-performance' and 'high-performance' applications. In fact, this division distinguishes the molecular (chemical) and crystalline (physical) components of graphene duality and indicates that if the use of the unique chemical properties of graphene at the molecular level does not cause resistance from its physics, opposite, the implementation of the unique physical properties is resisted by its chemistry.

The current paper concerns conflicts between the graphene chemistry and physics from the viewpoint of the molecular theory of graphene and suggests ways of overcoming the difficulties. Let us consider the above listed chemical properties in view of their influence on 'high-performance' applications

## 2. Radical character of the graphene substance and its morphological pattering

Molecular theory of graphene evidenced that chemical properties of graphene only slightly depend on the uniqueness of its regular one-atom-thick planar structure, while are mainly governed by peculiarities of electronic structure of 'sp$^2$-bonded carbon atoms that are densely packed in a honeycomb crystal'. The peculiarities are provided by odd electrons of benzenoid units with the smallest separations between them lying in the interval of 1.41-1.47Å. Both interval limits exceed the critical interatomic distance $R_{crit}$=1.395Å at which and below which the odd electrons are covalently bound and form the classical non-correlated π electrons [1-3]. Above $R_{crit}$ odd electrons become correlated, which causes their withdrawing from the covalent coupling and, consequently, radicalization of graphene. That is why any pristine graphene sample is an $N_D$-fold radical, where $N_D$ marks a total number of unpaired electrons. Regardless the size of the graphene sample, $N_D$ constitutes ~30% of the odd electron number $N_{odd} = N_{at} + N_{edge}$ [2, 3]. Here, $N_{at}$ and $N_{edge}$ are numbers of atoms in the entire sample and at its edges, respectively. Table 1 illustrates parameters of the odd electron correlation for a set of rectangular ($n_a$, $n_z$) nanographene (NGr) molecules with bare edges ($n_a$ and $n_z$ count the numbers of benzenoid units along armchair and zigzag edges, respectively). As seen from the table, the odd electrons of the molecules are quite strongly correlated.

Unpaired electrons are quantitative markers of chemical reactivity. Figure 1 presents the distribution of $N_D$ unpaired electrons over the atoms of the bare (15, 12) NGr molecule just exhibiting its 'chemical portrait'. As seen in the figure, the molecule circumference is the area of the highest chemical reactivity. The chemical reactivity of basal plane atoms is four times less in average. This basic edge property of graphene molecule well explains why morphological pattering of graphene sheets suggested for converting graphene from semimetal to semiconductors have failed [9].

Actually, cutting the sheets into nanoribbons increases the number of unpaired electrons drastically thus enhancing their radical properties. Inserting nanomeshes results in the same effect due to highly active periphery of the formed holes. Deposition of nanosize quantum dots highly disturbs the graphene substrate changing C-C bond length distribution and thus causing the $N_D$ growth if even not contributing by their own unpaired electrons. Therefore, cutting and drilling create a big 'edges problem' and do not seem to be proper technologies for the wished transformation of the graphene electronic system.

## 3. The collectivity of graphene electron system and chemical modification of graphene

Strong correlation of odd electrons provides a high level of their collectivization. For the first time it has been faced both experimentally [10, 11] and computationally [12] when performing nanolithography and atomic manipulation on the silicon Si(111)(7x7) surface by STM. As became clear later, the silicon surface, full of dangling bonds, presents an extensive reservoir of unpaired electrons [13, 14], which explains not only peculiarities of atomic manipulation on the surface but ensures its metallic conductivity and magnetic properties [13-15].

**Table 1**. Identifying criterion parameters of the odd electron correlation in rectangular nanographene molecules with bare edges [2]*

| Fragment $(n_a, n_z)$ | Odd electrons $N_{odd}$ | $\Delta E^{RU}$ ** kcal/mol | $\delta E^{RU}$ % *** | $N_D, e^-$ | $\delta N_D$, % ** | $\Delta \hat{S}_U^2$ |
|---|---|---|---|---|---|---|
| (5, 5) | 88 | 307 | 17 | 31 | 35 | 15.5 |
| (7, 7) | 150 | 376 | 15 | 52.6 | 35 | 26.3 |
| (9, 9) | 228 | 641 | 19 | 76.2 | 35 | 38.1 |
| (11, 10) | 296 | 760 | 19 | 94.5 | 32 | 47.24 |
| (11, 12) | 346 | 901 | 20 | 107.4 | 31 | 53.7 |
| (15, 12) | 456 | 1038 | 19 | 139 | 31 | 69.5 |

\* Criteria of odd electrons correlation

*Criterion 1*:
$$\Delta E^{RU} \geq 0, \text{ where } \Delta E^{RU} = E^R - E^U$$
presents a misalignment of energy. Here, $E^R$ and $E^U$ are total energies calculated by using restricted and unrestricted versions of the program in use.

*Criterion 2*:
$N_D \neq 0$, where $N_D$ is the total number of effectively unpaired electrons and is determined as
$$N_D = trD(r|r') \neq 0 \text{ and } N_D = \sum_A D_A .$$
Here, $D(r|r')$ and $D_A$ present the total and atom-fractioned spin density caused by the spin asymmetry due to the location of electrons with different spins in different spaces.

*Criterion 3*:
$$\Delta \hat{S}^2 \geq 0, \text{ where } \Delta \hat{S}^2 = \hat{S}_U^2 - S(S+1)$$
presents the misalignment of squared spin. Here, $\hat{S}_U^2$ is the squared spin calculated within the applied unrestricted technique while $S(S+1)$ presents the exact value of $\hat{S}^2$.

Criterion 1 follows from a well known fact that the electron correlation, if available, lowers the total energy. Criterion 2 highlights the fact that the electron correlation is accompanied with the appearance of effectively unpaired electrons that provide the molecule radicalization. Criterion 3 is the manifestation of the spin contamination of unrestricted single-determinant solutions; the stronger electron correlation, the bigger spin contamination of the studied spin state.

** AM1 version of UHF codes of CLUSTER-Z1. Presented energy values are rounded off to an integer

*** The percentage values are related to $\delta E^{RU} = \Delta E^{RU} / E^R(0)$ and $\delta N_D = N_D / N_{odd}$, respectively

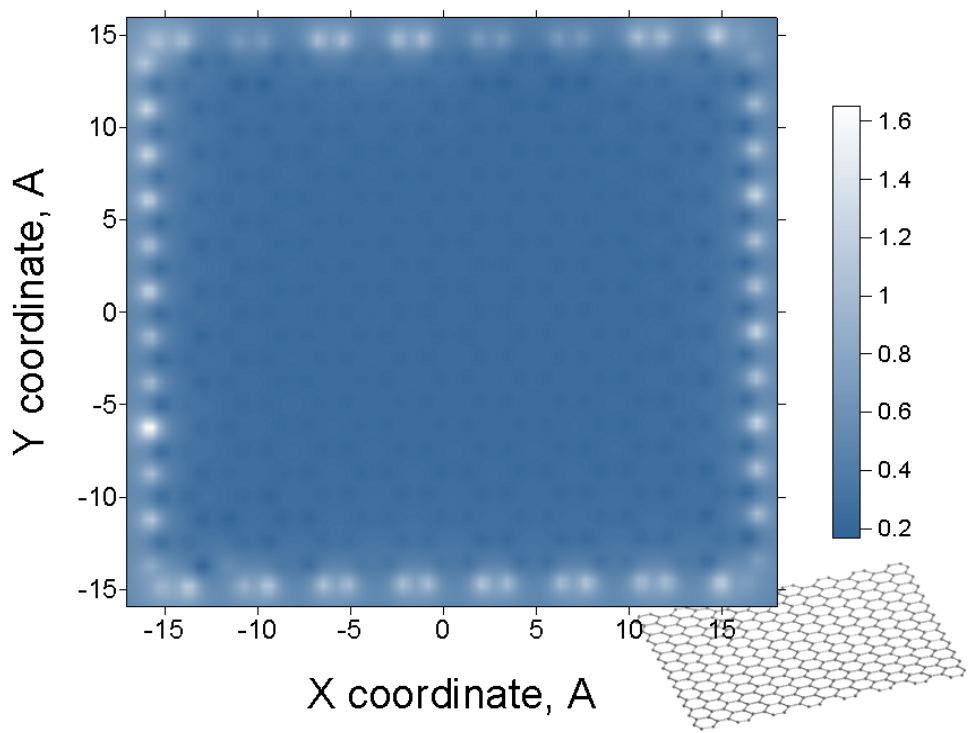

**Figure 1**. Atomic chemical susceptibility distribution over atoms of the (15, 12) NGr molecule with bare edges (UHF calculations, see details in [1]).

Despite of the fact that this effect is most strongly manifested in the case of silicon nanoobjects, both in silicon surface science and modern silicene projects, it is computationally more thoroughly studied in the case of carbon, in general [1], and graphene, in particular [2, 3]. A collective response of graphene unpaired electrons comes along with any external action whether it is magnetic or electric field, photoexcitation, mechanical loading, and chemical modification. However, when the first three are related to the object as a whole, due to which the collective character of response is quite expected, the latter two concern the object structure changes that might be quite local, particularly, in the case of chemical modification. This makes the latter a sensitive tool, revealing a microscopic image of the feature.

Changing in the internal structure of unpaired electrons may be traced by using the graphene molecule ACS maps, one of which related to the (15, 12) NGr bare molecule is shown in Fig. 1. As seen in the figure, in view of the activity of chemical modification, the molecule space is strictly divided into two parts covering circumference and basal plane, with an evident preference to the first one. Therefore, first steps of any chemical reaction occur at the molecule periphery. Since this reactivity area is largely spread in space, the formation of the first monoderivative does not inhibit the molecule reactivity so that the reaction will continue until the reaction ability is saturated. This means that any chemical modification of graphene is carried out as polyderivatization of the pristine molecule at its circumference. The molecular theory of graphene [2, 3] suggests a per-step-addition algorithm, governed by the highest ACS at each step, of the reaction description in due course of which it is possible to trace the unpaired electron behavior.

As turned out, already the first addition of any reactant, or modifier, to the edge atom of graphene molecule, chosen by the highest ACS, causes a considerable changing in the pristine ACS map thus allowing the exhibition of the second edge atom with the highest ACS to proceed with the chemical modification and so forth. This behavior is common to graphene molecules of any size and shape. In what follows, the behavior feature will be demonstrated on the example of the (5, 5) NGr molecule that was chosen to simplify the further presentation. Figure 2 presents a set of (5, 5) NGr polyhydrides and polyoxides obtained in the course of the first stage of the relevant per step reactions that concerns framing the bare molecule [16]. Two important conclusions follow from the figure. First, in spite of seemingly local change of the molecule structure caused by the addition, the second target carbon atoms does not correspond to the atom that is the second one of the highest activity in the ASC list of the pristine molecule. Second, this atom position as well as the sequence of steps varies depending on the chemical nature of the addends. Both two features are the result of the redistribution of C-C bond lengths over the molecule thus providing collective action of its unpaired electrons.

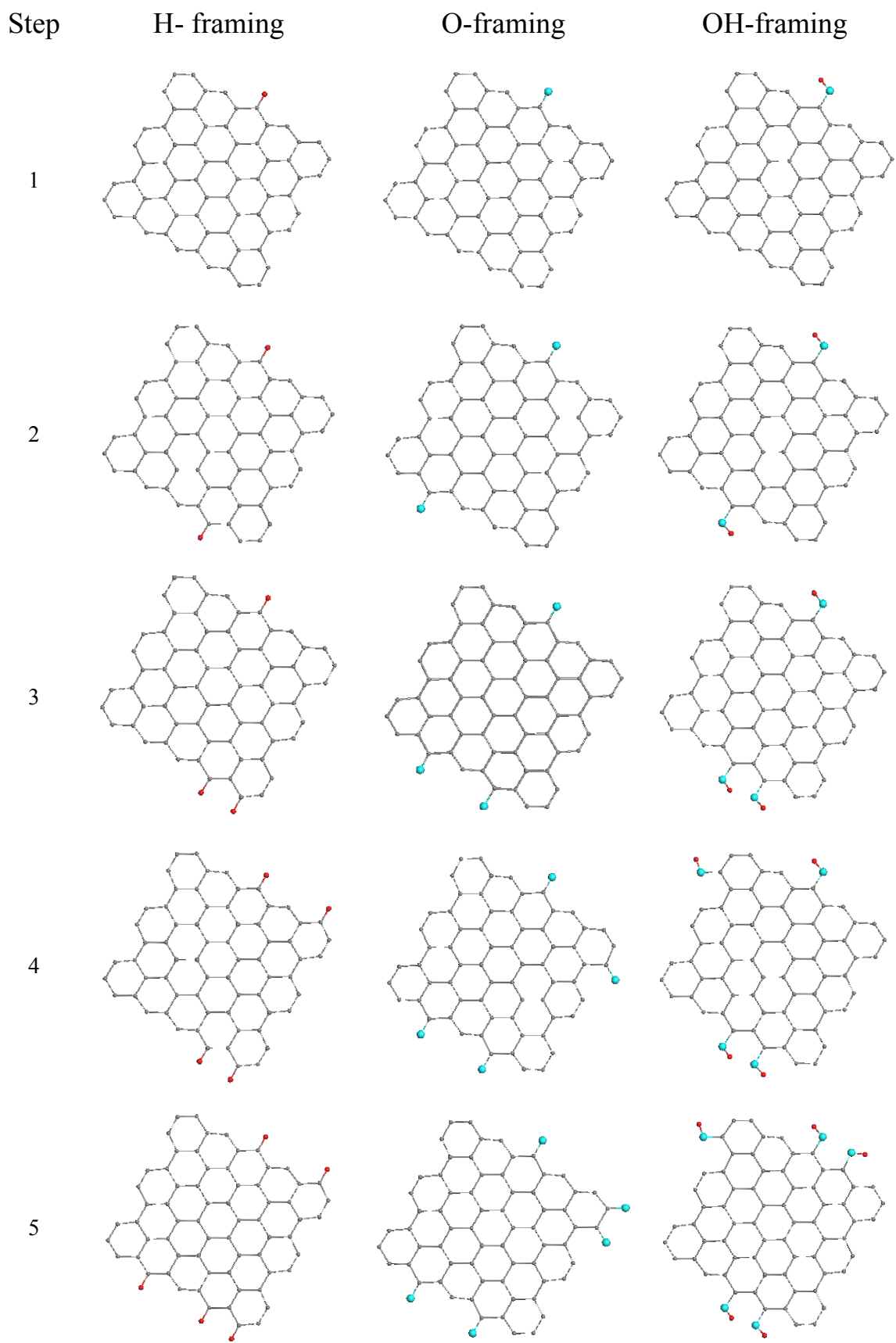

**Figure. 2**. Equilibrium structures of the (5,5) NGr polyhydrides and polyoxides related to $1^{st}$, $2^{nd}$, $3^{rd}$, $4^{th}$, and $5^{th}$ steps obtained in the course of the relevant stepwise reactions. Gray, blue, and red balls mark carbon, oxygen, and hydrogen atoms, respectively.

Chemical modification of graphene is not only a subject of interesting chemistry but has been repeatedly suggested as an efficient tool for the semimetal-semiconductor transferring [9] discussed in the previous sections. It should be noted that the suggestions are based on results of computational studies that concern drawn pictures of graphene fragments including those or other chemical modifiers artificially spread over graphene sheets (see, for example, [17, 18]). These and many other virtual structures, regularly distributed in space by applying periodic boundary conditions, exhibit electronic properties that are so badly needed for 'high-performance' electronics. However, the empirical reality is much less promising since so far none of regularly chemically modified graphene structure has been obtained. And collective behavior of graphene unpaired electrons is the main reason for the failure.

The wished regularly chemically modified graphene structures are related to the graphene polyderivatives that are formed with the participation of carbon atoms on the basal plane. However, as was shown earlier, reactions at the circumference precede those at the basal plane. Moreover, the latter cannot begin until the former are completed. In the predominant majority of the studied cases, the completion of the circumference reactions means the completion of the studied molecules framing. A thorough study of the circumference reactions has disclosed a very exciting feature: the framing of graphene molecules promotes the molecule cracking. Figure 3 present a set of ACS maps related to mono-hydrogen terminated ($H_1$-terminated below) NGr molecules of different size. The ACS maps of all the pristine molecules are of identical pattern characteristic for the (15, 12) MGr molecule shown in Fig. 1 just scaled according to the molecule size. As seen in the figure, the ACS maps of $H_1$-terminated polyderivatives show a peculiar division of the maps of the (15, 12) (3.275 x 2.957 $nm^2$) and (11, 11) (2.698 x 2.404 $nm^2$) NGr molecules into two parts in contrast to the maps of the (9, 9) (1.994 x 2.214 $nm^2$), (7, 7) (1.574 x 1.721 $nm^2$), and (5, 5) (1.121 x 1.219 $nm^2$) NGr molecules. The finding seems to demonstrate the ability of graphene molecules to be divided when their linear size exceeds 1-2 nm. The cracking of pristine graphene sheets in the course of chemical reaction, particularly, during oxidation, was observed many times. A peculiar size effect was studied for graphene oxidation in [19]. During 900 sec of continuous oxidation, micrometer graphene sheets were transformed into ~1 nm pieces of graphene oxide. Obviously, the tempo of cracking should depend on particular reaction conditions, including principal and service reactants, solvents, temperature, and so forth. Probably, under certain conditions, cracking can be avoided. Apparently, this may depend on particular conditions of the inhibition of edge atoms reactivity. However, its ability caused by the inner essence of the electron correlation is an imminent threat to the stability and integrity of the final product.

In some cases, the cracking is not observed when graphene samples present membranes fixed over their perimeter on solid substrates. Therewith, the reactivity of circumference atoms is inhibited and the basal plane is the main battlefield for the chemical modification. Still, as in the case of circumference reactions considered earlier, the highest ACS retains its role as a pointer of the target carbon atom for the subsequent reaction steps. However, the situation is much more complicated from the structural aspect viewpoint. Addition of any modifier to the carbon atom on the basal plane is accompanied by the $sp^2 \rightarrow sp^3$ transformation of the valence electrons hybridization so that for regularly packed chemical derivatives, the benzenoid carcass of pristine graphene should be substituted by the cyclohexanoid one related to the formed polyderivatives. When benzene molecules and, subsequently, benzenoid units are monomorphic, cyclohexanes, and thus cyclohexanoid units, are highly heteromorphic. Not very big difference in the isomorphs free energy allows for coexisting cyclohexanoids of different structure thus making the formation of a regular structure a rare event. Actually, the regular crystalline-like structure of a graphene polyhydride, known as graphane, was obtained experimentally when hydrogenating fixed graphene membranes accessible to hydrogen atoms from both sides [20]. In the same experiment, fixed membranes accessible to hydrogen atoms from one side only showed irregular amorphous-like structure. The empirical findings were supported by computations based on the consideration of stepwise hydrogenation of fixed and free standing membranes accessible to hydrogen atom from either two or one side [21]. Figure 4 shows equilibrium structure of (5, 5) NGr membranes, edge atoms of which were terminated by two hydrogen atoms that ensured a complete inhibition of chemical activity at the membrane circumference and which were hydrogenated to the saturation limit. The figure clearly shows how much depends the final product structure on the conditions applied to membranes. Thus, hydrogenation of fixed membrane accessible to hydrogen atoms from both sides results in the formation of a perfect regular structure provided by chair-like cyclohexanoids (Fig. 4a). When the membrane is accessible from one side only, it looks like a canopy, the structure of which is mainly formed by table-like cyclohexanoid isomorphs (Fig. 4b). These two configurations were obtained in the experiment [20]. Unlatching the membrane around the perimeter drastically changes the structural configuration of the final products. In the case of two-side accessibility of the membrane, the regular graphane structure is transformed into a mixture of quasi regular areas that cover chair-like and boat-like formed hydrides (Fig. 4c). In the case of one-side accessibility, the free standing membrane is rolled into a basket (Fig. 4d).

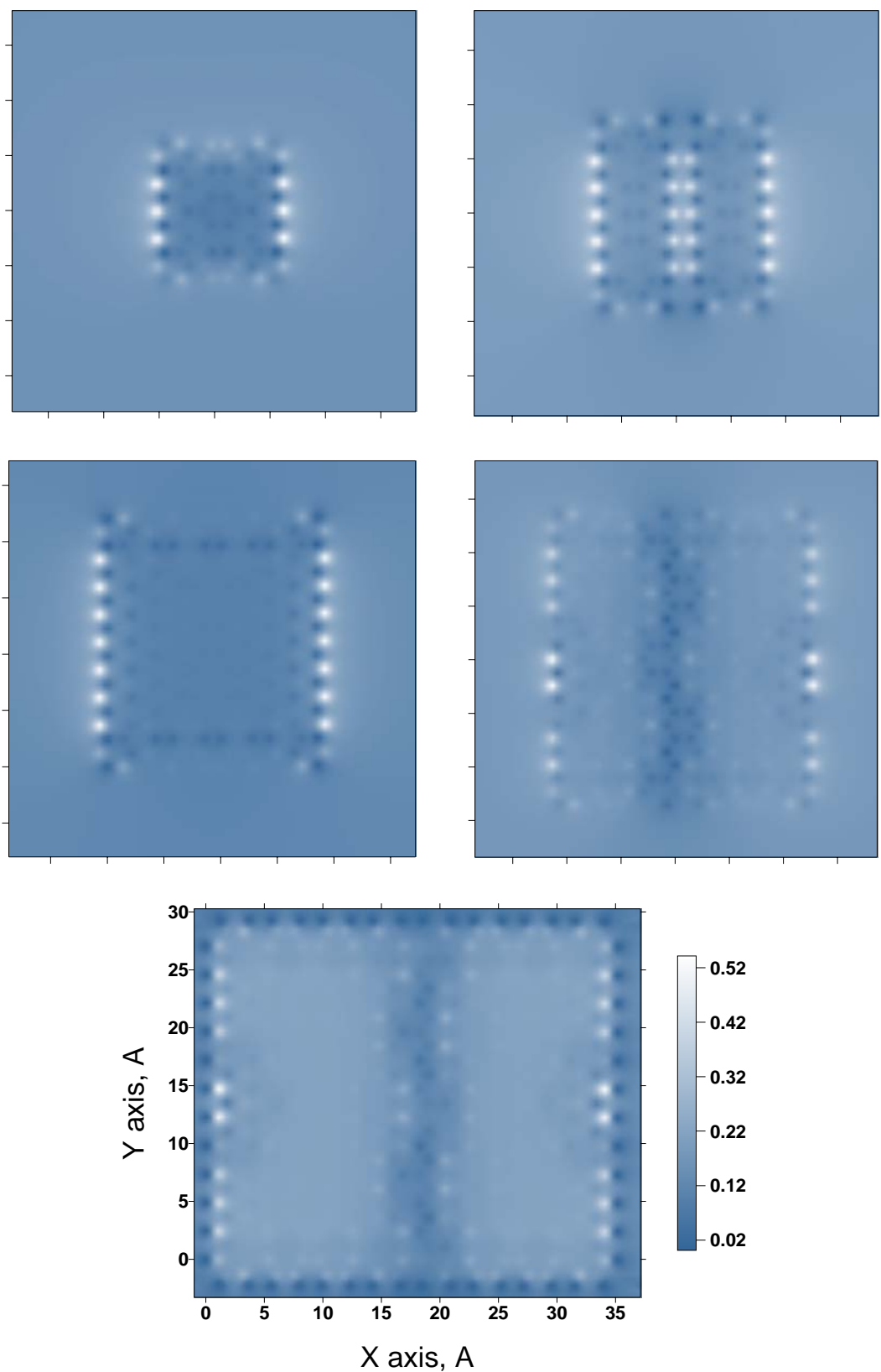

**Figure. 3**. Atomic chemical susceptibility distribution over atoms of the (5, 5), (7, 7), (9, 9), (11, 11), and (15, 12) NGr molecules with $H_1$-terminated edges (UHF calculations). All the images are given in the same space and ACS scales shown on the bottom.

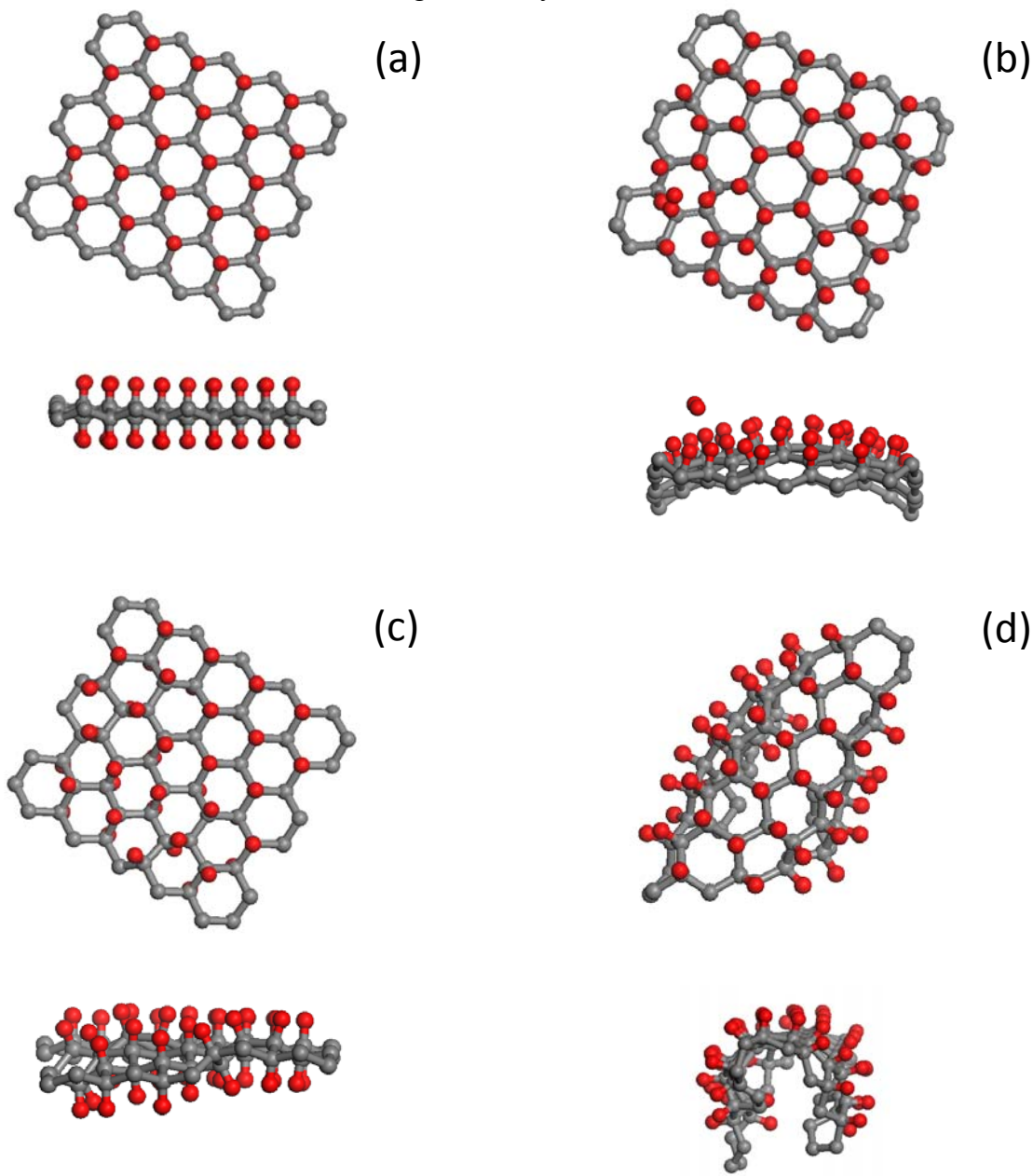

**Figure 4.** Top and side views of the equilibrium structure of the hydrogen-saturated (5, 5) NGr hydrides. The pristine (5, 5) NGr membranes are either fixed (a, b) or free standing (c, d) and accessible to hydrogen atoms from both (a, c) and one (b, d) side.

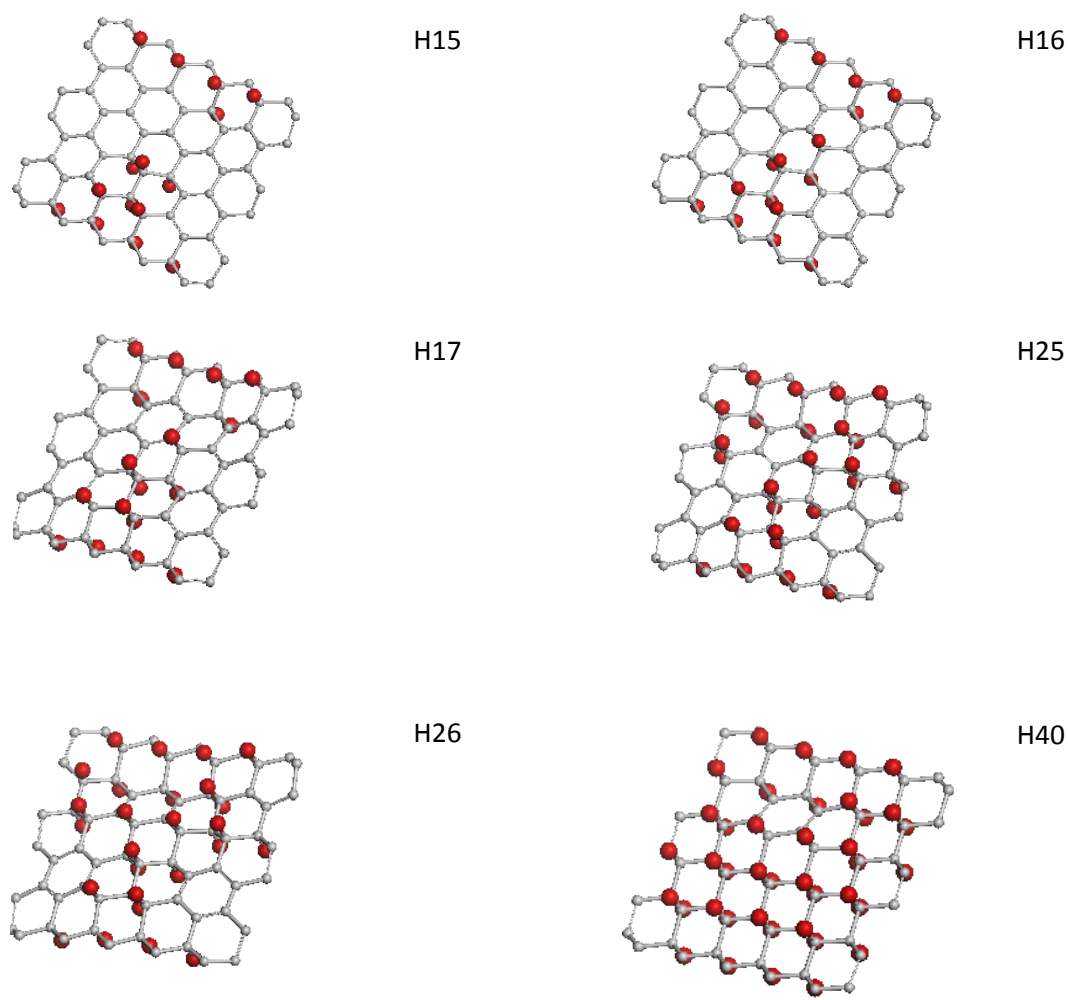

**Figure 5.** Equilibrium structures of (5, 5) NGr hydrides at two-side accessibility of the fixed memrane to hydrogenation. The framing hydrogen atoms are not shown for simplicity. Figures show the number of hydrogen atoms attached in the basal plane.

As shown above, it is possible to proceed with chemical modification of graphene within the basal plane only after a complete inhibition of high chemical activity of atoms at the circumference. In spite of the ACS values within the area are much less than at a bare circumference, they still constitute ~0.3-0.1$e$ that is quite enough to maintain active chemical modification. However, the reality drastically differs from the wished chemical pattering of graphene sheets whose virtual image present the final product of the pattering as regular carpets similar to flowerbeds of the French parks. The reality is more severe and closer to designs characteristic of the English parks. The matter is that the collective of unpaired electrons, which strictly controls the chemical process at each step, has no means by which to predict the modifier deposition sites many steps forward. And it is clear why. Each event of the modifier deposition causes an unavoidable structure deformation due to local $sp^2 \rightarrow sp^3$ transformation in the place of its contact with graphene. The relaxation of the deformation, as was seen in Fig. 2, extends over a large area, which, in turn, is accompanied by the redistribution of C-C bond lengths. Trying to construct a pattering, it is impossible, while not making calculations, to guess at what exactly carbon atom will concentrate the maximum reactivity, highlighting the latter as a target atom to the next deposition. Therefore, even two simultaneous depositions cannot be predicted, not to mention such complex as quantum dots or nanoribbons. That is why a wished regular chemical pattering of graphene basal plane exists only on beautiful pictures. The real situation was studied in detail in the case of graphene hydrogenation [22], exhibiting the gradual filling of the basal plane with hydrogen at random. Figure 5 presents the calculation view of the (5, 5) NGr membrane top filling with atomic hydrogen related to 34 to 91% coverage [21] that is in full consistence with the pictures observed in [22].

Final products of the addition reactions on basal planes of graphene strongly depend on the addends in use. Thus, in the case when the latter are either mono-atomic (hydrogen, fluorine, oxygen) or two-atomic (hydroxyl), the final products shown in Fig. 6 and related to the completing of the reactions at two-side accessibility of the (5, 5) NGr fixed membrane to the addends, are highly disordered beside the only one related to graphane. None of the regular motives was observed in all the cases in the course of step wise reactions.

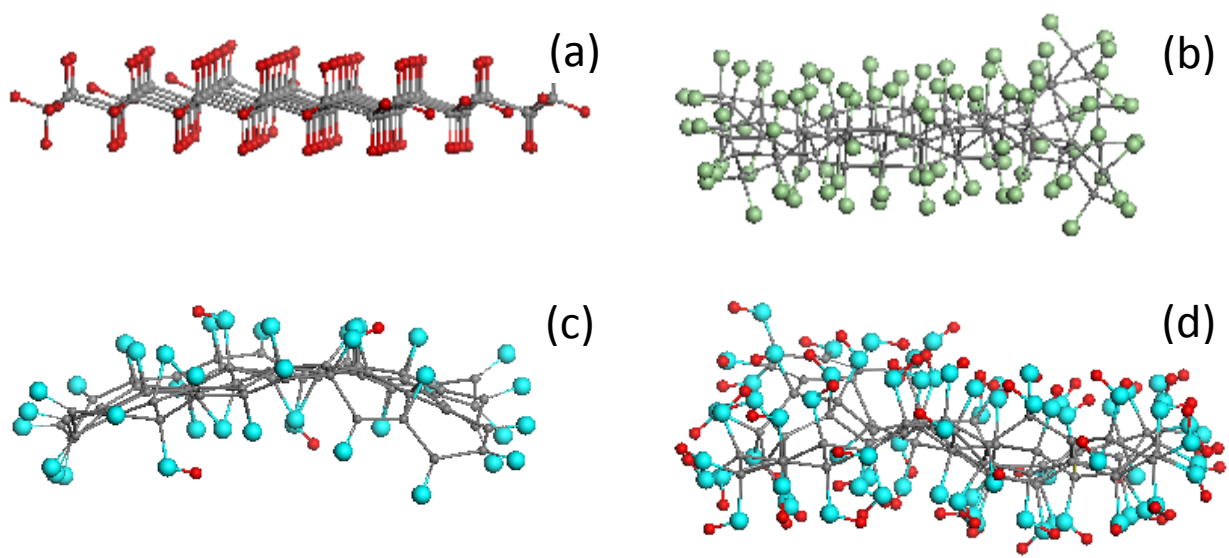

**Figure. 6**. Equilibrium structures of the addend-saturated (5,5) NGr polyhydride (a), polyfluoride (b), epoxy– and hydroxy-oxides (c and в, respectively). Computational stepwise synthesis (UHF calculations).

4. **The collectivity of electron system and mechano-chemical reactions of graphene**

The molecular theory considers mechanical behavior of graphene from the position of polymer science [23, 24]. The deformation and rupture of polymers have long been successfully treated from the standpoint of mechano-chemical reactions [25]. The approach allows for exhibiting the chemical uniqueness of graphene in view of its mechanical behavior.

Graphene deformation and rupture concern stretching and breaking of C-C bonds. Since these bond lengths determine ACS of the involved carbon atoms, the bond stretching obviously causes the enhancement of the atoms ACS until the latter reaches maximum at the bond breaking. For the first time, the effect was carefully studied computationally on the example of uniaxial tension of the (5, 5) NGr molecules, both bare and mono- and di-hydrogen terminated [23, 24]. Figures 7-9 in the upper parts exhibit the equilibrium structures of the molecules in their nondeformed and broken final states at two deformational modes attributed to the tension along (zigzag mode) and normally (armchair mode) to the C-C bond chains. Two lower parts present stress-versus-strain dependences and changing of the molecular chemical susceptibility (MCS) of the molecules expressed as the total number of unpaired electrons $N_D$ [1-3]. In all the cases, the molecule extension was carried out via a simultaneous elongation of a number of mechanical internal coordinates, particularly inserted (see the coordinate description in [23]), at a constant pitch of 0.1 Å. A drastic difference in the zigzag and armchair modes behavior, characteristic for the bare molecule, is gradually smoothed when going to mono- and di-hydrogen terminated molecules. Taking together, the picture of successive stretching and breaking of the molecule C-C bonds, presented in Figs. 7-9, exhibits a peculiar topochemical character of graphene mechano-chemical reactions that is common to all graphene samples of any size and shape [24]. Actually, the extended zigzag deformation, which is accompanied with the formation of a long one-atom chain (Fig. 7), was observed in a study [26] that implemented zigzag deformation mode of bare graphene ribbons in practice.

A saw-tooth pattern is a characteristic element of all the MCS dependences shown in Fig.7-9. The pattern clearly discloses the stretching-and-breaking event related to one or a few bonds. Accordingly, the difference of the MCS behavior of different molecules at two modes can be described as different numbers or stretching-and-breaking events occurred until the final breaking of the molecules. As for each individual event, the stretching phase is accompanied with the MCS growth, while at the breaking stage the MCS can either grow or decrease depending on which namely fragments are formed after the breaking and what are the C-C bond lengths of the equilibrium structures.

Each stretching-and-breaking event is ended with the formation of this or that defect so that non-defect (non-damage) deformation of graphene can be realized at the initial stage of deformation only. Just this situation has been recently implemented in practice [27] where a convincing evidence of the enhancement of chemical reactivity of graphene, subjected to tensile deformation, was obtained.

The discussed above concerns MCS determined by the total number of unpaired electrons. For details of chemical interaction of the strained graphene with different modifiers, we have to address to the ACS of the graphene atoms. Figures 10 and 11 present the ACS maps of strained bare and mono-hydrogen terminated (5, 5) NGr molecules at different stages of deformation for two deformation modes. Similarly to conventional chemical modification, such maps control the filling of basal plane of strained graphene with modifiers. As seen in the figure, the first filling event will occur at different atoms of the molecules depending on both the termination of edge atoms and deformational mode, or, by other word, the direction of mechanical loading. Obviously, the filling itinerary cannot be predicted as was in the case of chemical modification of graphene discussed earlier. But it can be traced in the course of a stepwise computational modification of strained graphene. Due to endless possibilities concerning graphene samples of different size and shape, different directions of mechanical loading, different modifiers, and so forth, this way to model chemical modification of strained graphene precisely does not look promising. However, a certain set of particularly designed calculations can evidently allow disclosing common features of this complicated event. Empirically, so far we can rely upon the saturation regimes giving post-factum these or those products. The only thing is definitely clear that mechanical loading considerably enhances graphene chemical modification and may allow chemical reactions that would otherwise not proceed.

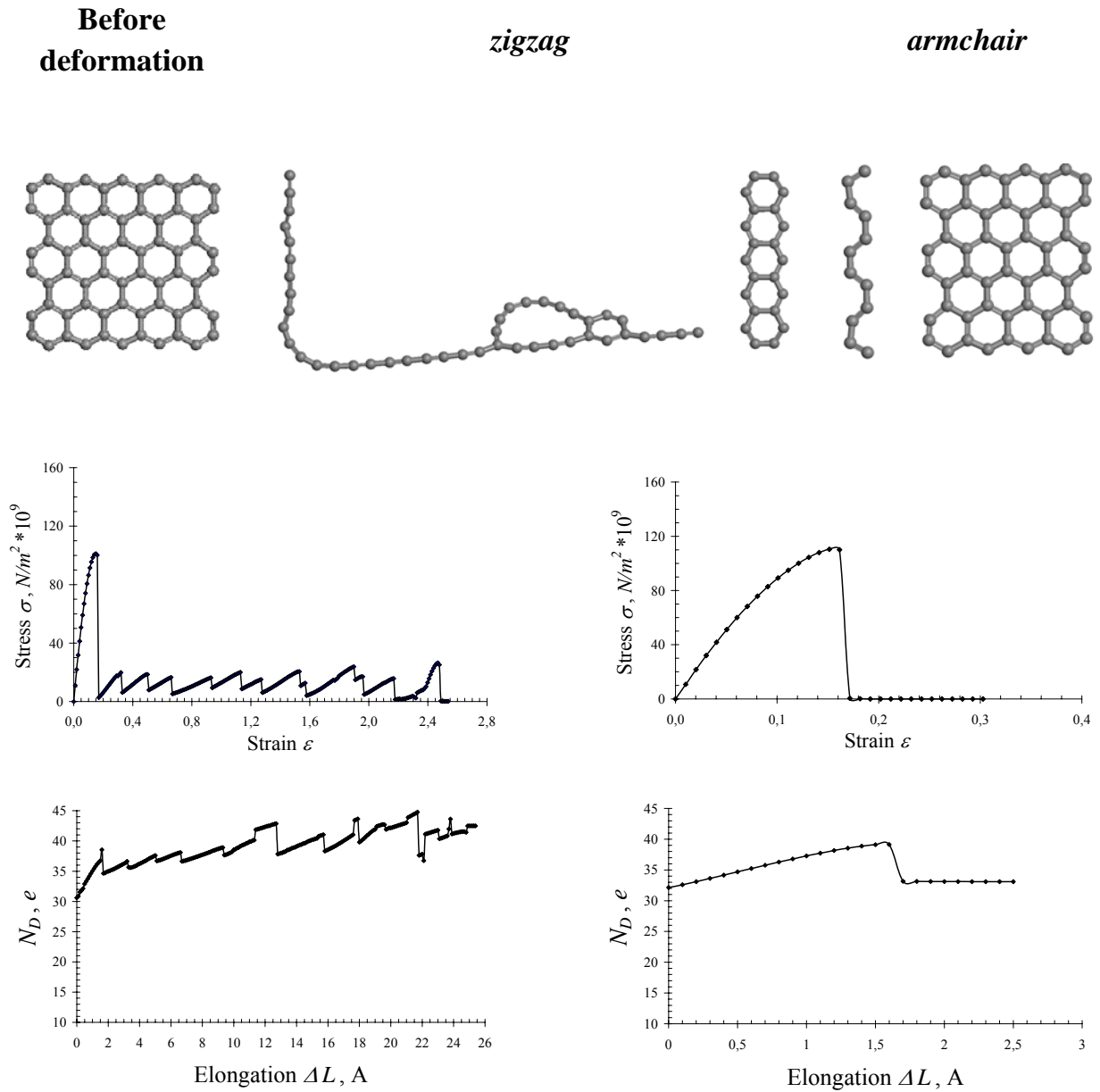

**Figure 7**. Equilibrium structures of the (5,5) NGr molecule with bare edges before and after completing tensile deformation in two modes of deformation (top); Stress-versus-strain dependences (middle); Evolution of the odd electrons correlation in terms of the total numbers of effectively unpaired electrons (bottom).

The considered deformation caused by the application of external stress, can be attributed to a dynamic one. Besides, a number of static deformation modes exist. In reality, these modes are related to the deformation of the carbon skeleton in the places of graphene sheet roughnesses of different origin, such as wrinkles and bubbles that are formed when graphene sheets are placed on some substrates. When the deformation causes stretching of the sheet skeleton, it is mandatory accompanied with enhancing the chemical reactivity. The effect can be modeled by the chemically stimulated stretching of the skeleton that can be traced by comparing those related to hydrides of the (5, 5) NGr molecule of the canopy-like and basket-like ones obtained in the course of one-side hydrogenation of either fixed or free standing membrane, respectively (the hydride structure see in Fig. 3). As shown [2, 3], the skeleton deformation causes increasing of the total number of unpaired electrons $N_D$(MCS) form 31$e$ for a bare molecule to 46 $e$ and 54 $e$, for the canopy-like and basket-like skeletons, respectively. Figure 12 shows the distribution of these unpaired electrons over the skeleton atoms of non-deformed and strained molecules. As seen in the figure, the brightness of the skeleton electron-density image greatly increases when the skeleton curvature becomes larger (draw attention on a large vertical scale of plottings presented in the top figure).

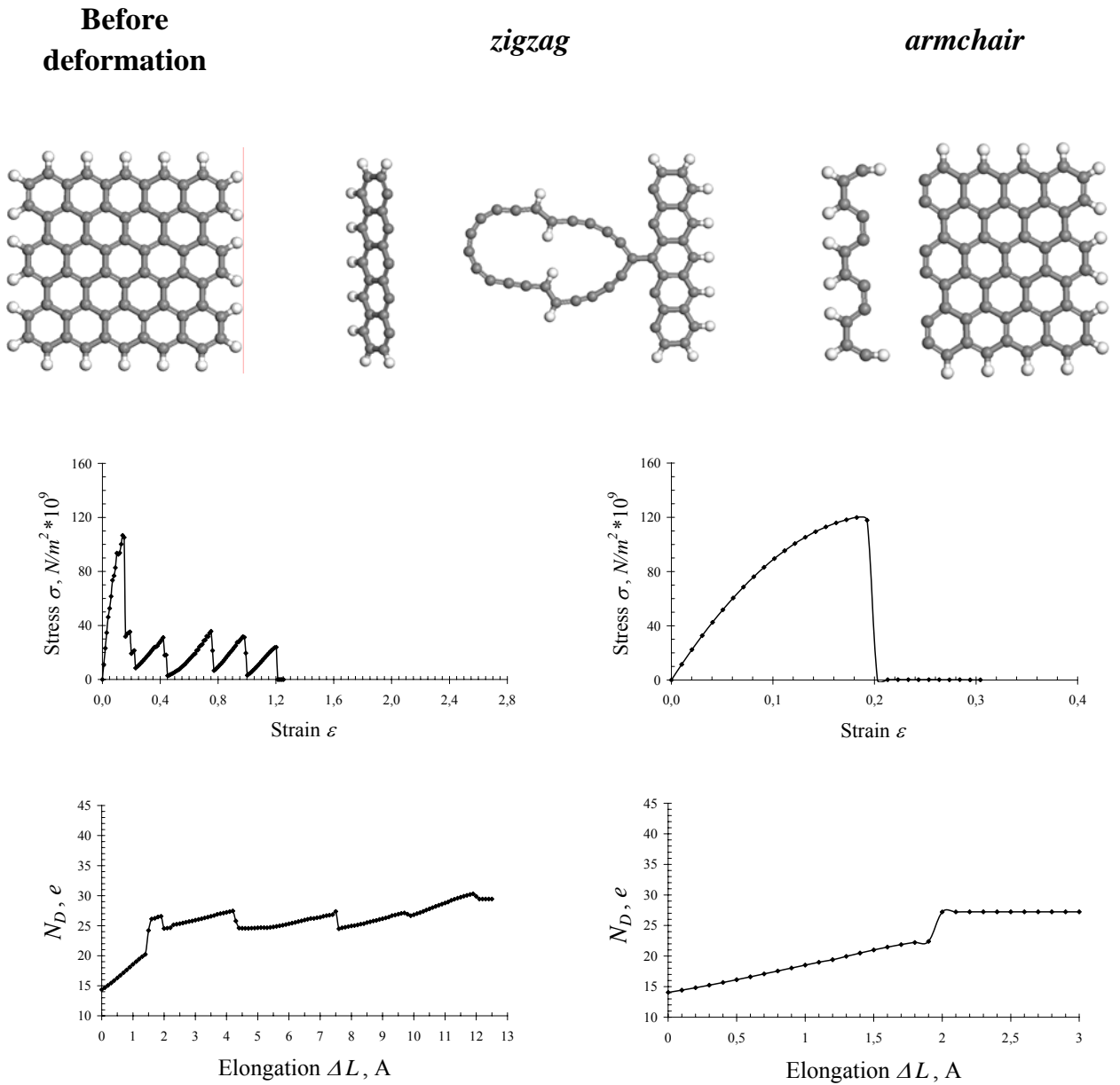

**Figure 8**. Equilibrium structures of the (5,5) NGr molecule with $H_1$-terminated edges before and after completing tensile deformation in two modes of deformation (top); Stress-versus-strain dependences (middle); Evolution of the odd electrons correlation in terms of the total numbers of effectively unpaired electrons (bottom).

The deformation-stimulated MCS rise leads to a number of peculiar experimental observations. As follows from Fig.12, if observed by HRTEM, the basket-like skeleton might have look much brighter than the canopy-like one and especially than the least bright pristine molecule. In view of the finding, it is naturally to suggest that raised above the substrate and deformed areas of graphene in the form of bubbles, found in a variety of shapes on different substrates [28, 20], reveal peculiar electron-density properties just due to the stretching deformation. Small (5, 5) NGr molecule presented in Fig. 12 cannot pretend to simulate the picture observed for micron bubbles, but it exhibits the general trend that takes place in bubbles, as well. This explanation of a particular brightness of the bubbles at HRTEM images looks more natural than that proposed from the position of an artificial 'gigantic pseudo-magnetic field' [28].

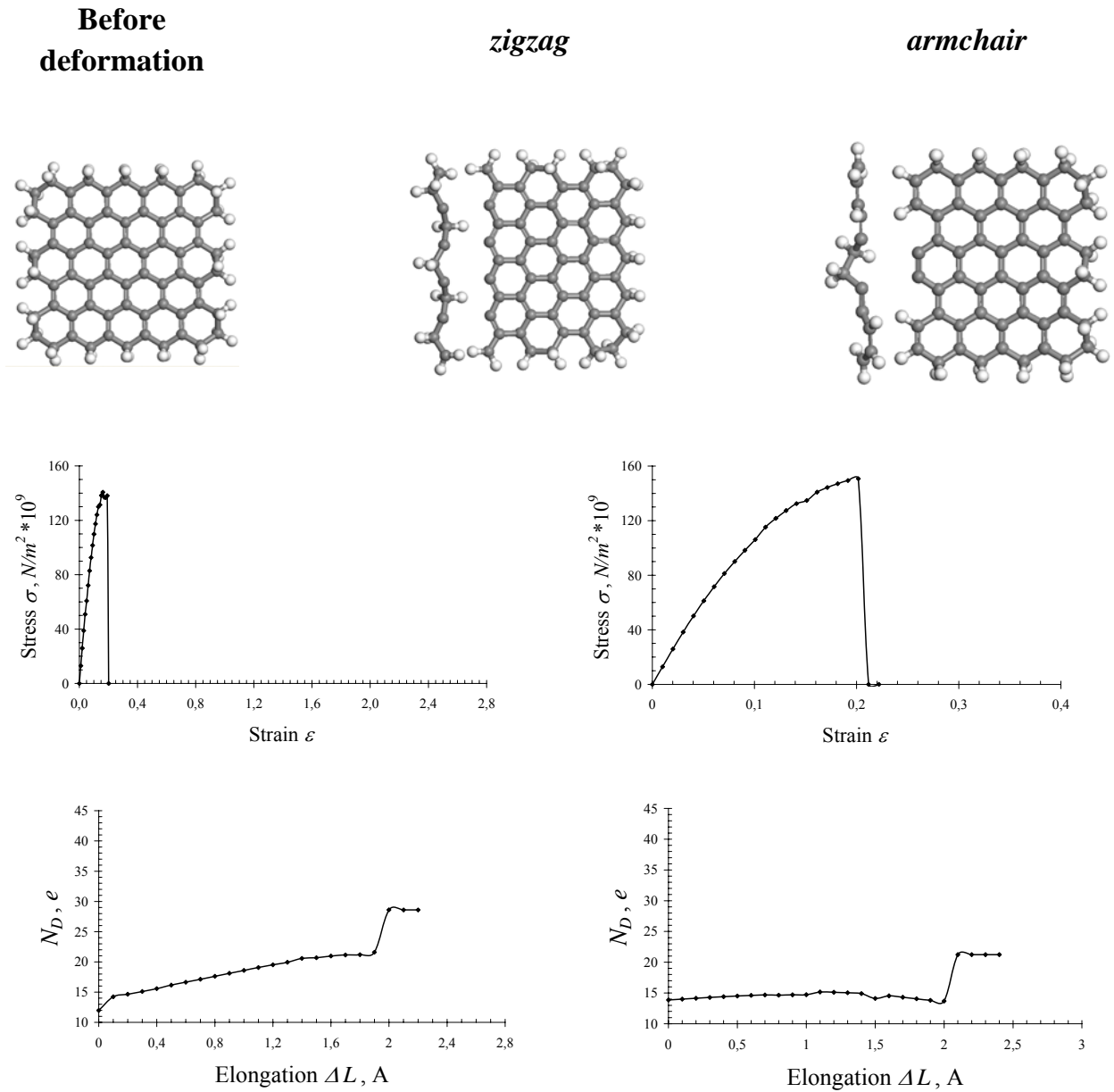

**Figure 9**. Equilibrium structures of the (5,5) NGr molecule with $H_2$-terminated edges before and after completing tensile deformation in two modes of deformation (top); Stress-versus-strain dependences (middle); Evolution of the odd electrons correlation in terms of the total numbers of effectively unpaired electrons (bottom).

The next observation concerns high-density wrinkles formed at a monolayer graphene structure grown on Pt(111) [30]. As shown, the wrinkles can act as nanosized gas-inlets in the graphene oxidation due to enhanced reactivity of wrinkles to oxygen. Analogous effect of enhanced reactivity was observed for monolayer graphene deposited on a Si wafer substrate decorated with $SiO_2$ nanoparticles (NPs) and then exposed to aryl radicals [31]. As shown, the aryl radicals selectively react with the regions of graphene that covered the NPs thus revealing the enhanced chemical reactivity of the deformed graphene spots.

*Armchair deformational mode*

| 0 | 1.00 | 1.09 | 15 |
| 10 | 1.07 | 1.09 | 16 |
| 13 | 1.08 | 1.0 | 17 |

*Zigzag deformational mode*

| 0 | 1.00 | 0.89 | 15 |
| 10 | 0.94 | 0.94 | 16 |
| 13 | 0.91 | 1.32 | 17 |

**Figure 10.** Image ACS maps and equilibrium structures of the (5, 5) NGr molecule with bare edges in the course of the first stage armchair-mode (top) and zigzag-mode (bottom) tensile deformation. Black and white figures number steps and equalizing coefficients, respectively. The same intensity scale is related to all the maps.

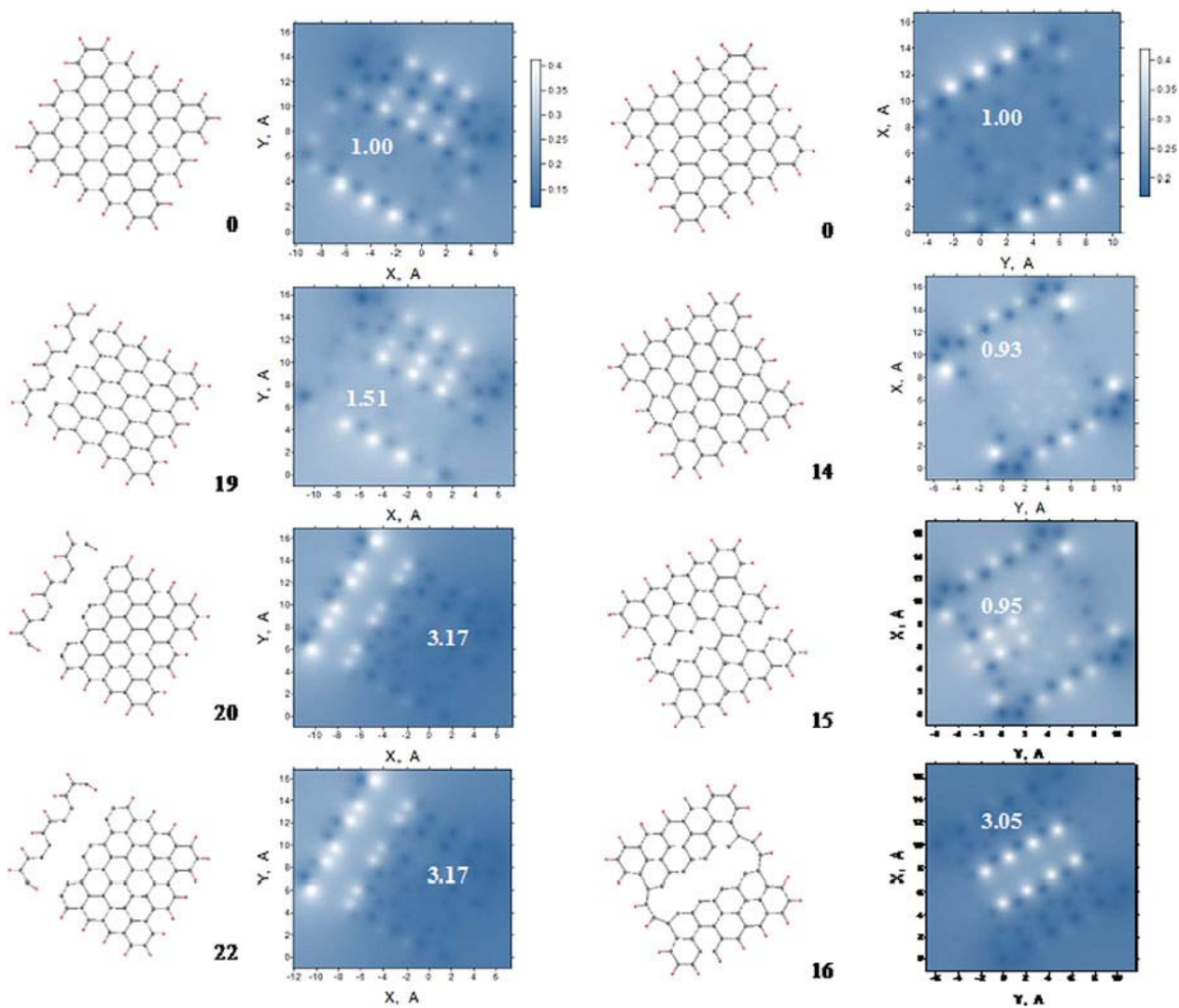

**Figure 11**. Image ACS maps and equilibrium structures of the singlet-hydrogen-terminated (5, 5) NGr molecule with $H_1$-terminated edges in the course of the first stage of the armchair (left) and zigzag (right) -mode tensile deformation. Black and white figures number steps and equalizing coefficients, respectively. Scales are related to all the maps within the deformational mode.

Beside the effects considered in the current section, the deformation-stimulated rise of the total number of unpaired electrons, which evidences the strengthening of the odd electron correlation, results in a possible activation of the sample magnetization. The molecular aspect of the graphene magnetism and the deformation effect are considered elsewhere [2, 3].

## 5. Graphene deposition on substrate inhibits the odd electron correlation

The chemical variability of graphene provides a large range of its application at molecular level in such devices as touch screens, rollable e-paper, foldable OLED, varied sensors, solar cells, and many others. All the community forms the group of 'low-performance' applications [8]. At the same time, this uniqueness makes graphene extremely sensitive to any external action, say, unavoidable deformation, thus preventing its application to 'high-performance' electronics. Obviously, to reach stability it is necessary to inhibit the graphene correlation status thus inhibiting its chemical radicalization. Deposition of graphene monolayers on appropriate substrate is seen as a very promising way. Actually, the interaction with substrate atoms inhibits radicalization while conserves 2D solid electronic properties. As shown recently [32], the electron energy of graphene monolayer, deposited on silicon carbide, boron nitride, and quartz surfaces remains $k$-linearly dependent at small $k$, albeit of different slope. If to substitute flat substrates with regularly decorated by using not rarely distributed NPs as in [31], but regular NPs grids, there is an amazing opportunity to exercise tuning the electronic gap of graphene by its covering over the NPs grids of different design. Besides, this hilly graphene scarf can serve as a template for a particular 'antilithography' presenting a hilly grid of chemically reactive spots to be subjected to a wished chemical modification.

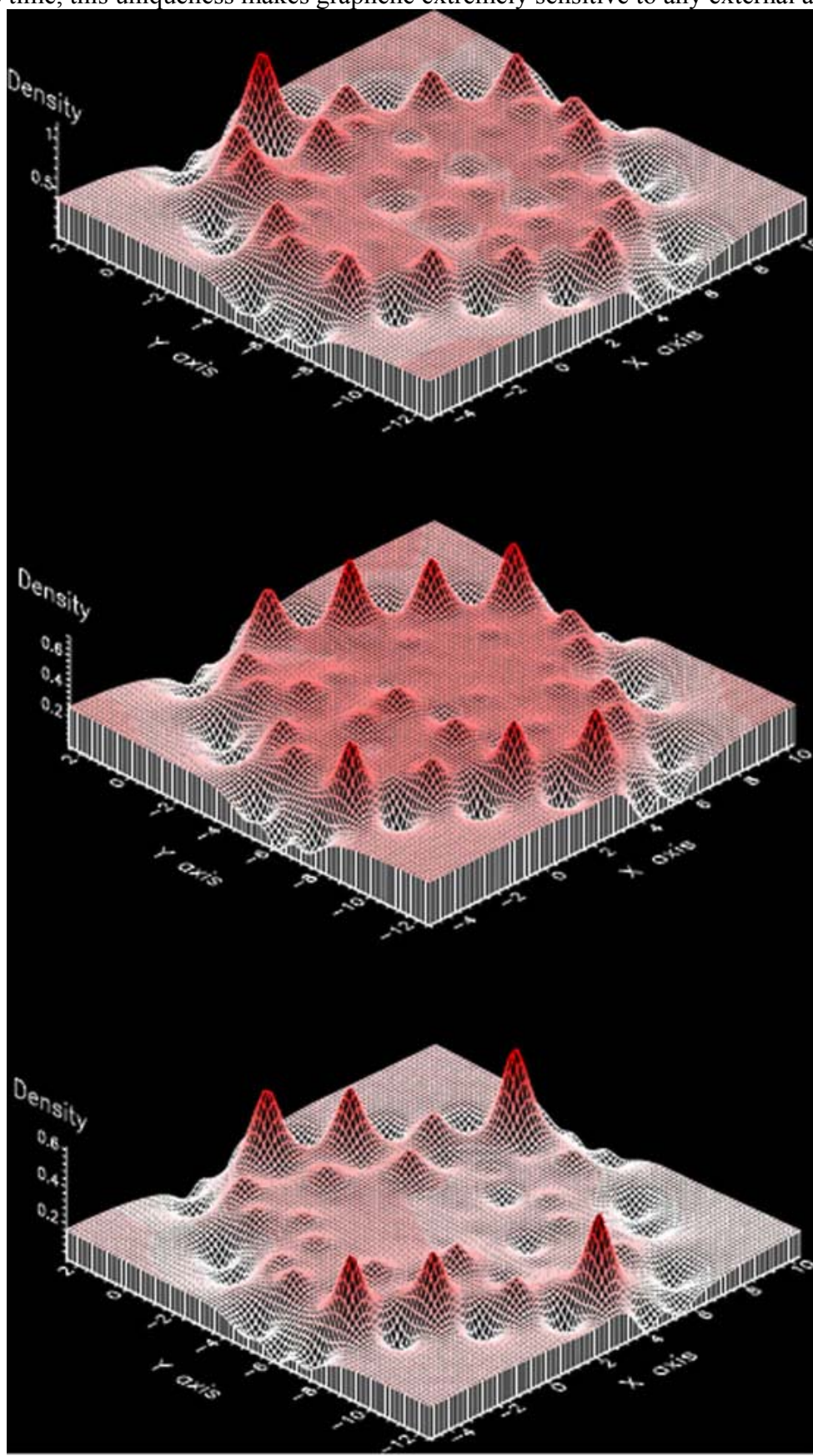

**Figure 12.** Effectively-unpaired-electron-density images of carbon skeletons of the pristine (5, 5) NGr (a), canopy-like (b) and basket-like (c) fixed membranes.

Monoatomic adsorbed carbon layers of hexagon pattering can be considered as the extreme case of graphene-substrate composites. Not all substrates admit the hexagon patterns for the adsorbate but a high experience of surface science allows for seeking the best partners. The successful execution of this trend is observed for the silicon analogue of graphene – silicene. While free standing silicene, once virtual substance, the object of computational study only, cannot exist [1, 33, 34], silicene-substrate composite as the monolayer of adsorbed silicon over Ag(111) surface is the subject of physical reality, demonstrating the 2D feature of the electronic spectrum [35]. At the same time, as reviewed in [36], Si monolayer on Ag(111) surface showed a strong resistance to oxidation, approximately $10^4$ times higher than a clean Si(111)-(7x7) surface. Figure 13 presents a comparative view of the ACS distribution over the atoms of the unit cell of the Si(111)-(7x7) surface and the (3, 7) nanosilicene molecule. Once differing by the total number of unpaired electrons (~93 $e$ and ~76 $e$ for the surface and molecule, respectively), the amplitude analysis shows comparable ACS values for the molecule and first three layers of the surface evidencing similar chemical reactivity of the species. Four-orders-of-magnitude inhibition of the molecule reactivity tells about its strong chemical coupling with the substrate. In spite of this, the coupling does not prevent from exhibiting peculiar 2D characteristics of the electronic spectrum [35]. A further chemical modification of the layer, aimed at tuning the energy gap, is greatly facilitated and made local due to the absence of odd electrons.

## 6. Conclusion

Graphene is extremely interesting object for scientific investigations. It holds many secrets, the disclosure of which has already caused and will continue to cause the appearance of more new concepts related not only to its physical and chemical properties but to physics and chemistry as such. In view of graphene applications, the chemical and physical concepts are tightly interrelated thus either enhancing each other or conflicting. The current paper presents a quantitative explanation of the conflict between high chemical reactivity of graphene and its stability needed for 'high-performance' physical application. A fragile but strong dependence of the graphene C-C bond structure on the collective behavior of its odd electrons lays the foundation of the conflict. Easily responding to any external action, C-C bond net is redistributed thus causing considerable, sometime even drastic, changes in the atomic chemical susceptibility of the graphene atoms, resulting in its instability and preventing the development of reliable technologies of tuning graphene properties by external actions. One of the ways to achieve stability is to inhibit graphene radicalization, by other words, to press the correlation of its odd electrons. The inhibition can be performed by placing graphene on a flat substrate that may provide the interaction between graphene and substrate atoms. The remaining hexagon pattering will ensure peculiarities of the 2D electronic spectrum. The underneath substrate surface may be artificially configured with nanostructured grids of additionally deposited units, whereby the formed hilly graphene scarp may be used both *per se* and as a template for further regular

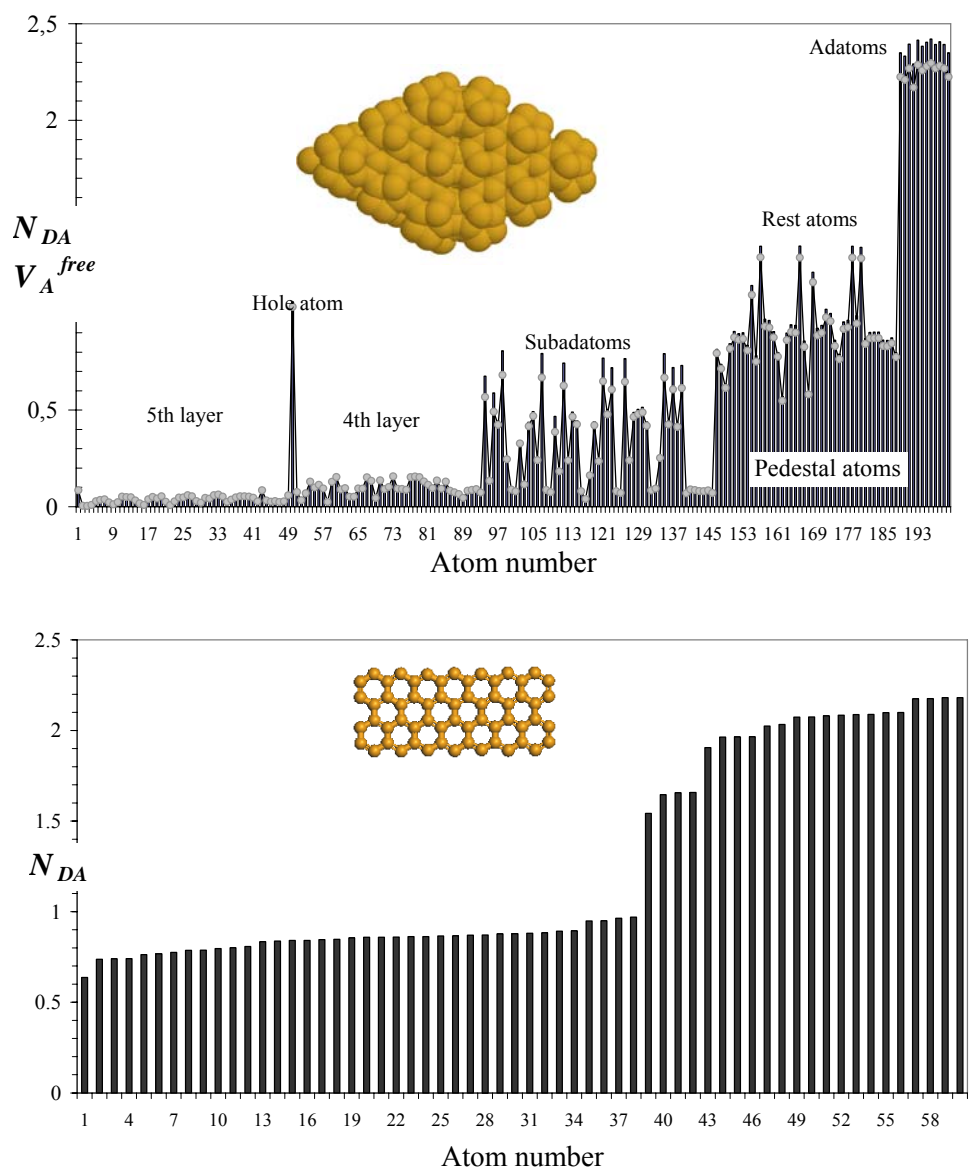

**Figure 13.** Atomic chemical susceptibility (histograms) and free valence (gray spots) distribution over atoms of the silicon Si(111)(7x7) surface unit cell (top) and (3, 7) nanosilicene molecule (bottom). Inserts are equilibrium structures of the objects,

chemical modification to tune the electronic properties in a wished manner. This line of research is difficult to be performed, both experimentally and computationally. However, there is every reason to expect the emergence of new conceptual results along the way.